# Inventions on Tree Navigators used in Graphical User Interface

**Umakant Mishra**

Bangalore, India

http://umakantm.blogspot.in

**Contents**

## 1. Introduction

A tree view or tree navigator is used to display hierarchical data organized in the form of a tree. In a tree structure there are parent and child nodes. The child nodes may further have descendants to n levels. One of the popular example of tree navigator is the file explorer in Microsoft Windows.

A conventional tree navigator works well when the numbers of nodes are small. The problem starts with the size of the tree navigator. To navigate a thousand nodes in multiple hierarchies, the user has to do a lot of expansion, contraction, scrolling and many difficult exercises.

There are many methods to make the navigation easy. Some of these are expanding and collapsing branches, splitting the tree, displaying a parent node in a separate tree, zooming branches, scrolling in various directions etc. It is still a difficult exercise to handle large trees efficiently. The effort still continues to manage large number of nodes with faster speed, greater control, user friendliness and aesthetics.

This article illustrates five inventions on tree navigators selected from US patent database. Each of them tries to solve various problems relating to the tree navigator in different ways. Each invention is also analyzed from a TRIZ perspective.

## 2. Inventions on tree navigators

### 2.1 Navigating applications and objects in a GUI (6078327)

**Background problem**

A tree navigator is used in a GUI to display hierarchical data in a tree structure. This structure is used in File Manager (Win 3.x) and Explorer (Win 95 and later) to display the contents of disk drives, directories, sub-directories and files. However, these utilities can only be used with the data storage devices and file system directories provided by the computer. There is a need for a utility that allows user to display and traverse the hierarchical structure of applications or documents and their elements.

**Solution provided by the invention**

Patent 6078327 (invented by Liman et al., assigned by IBM, issued Jun 2000) discloses a method of using a navigator window to display and traverse application programs. The navigator window comprises of a collapse-expand tree control for traversing one or more applications created by the application program. Each of the applications is comprised of one or more objects and the application program window displays an application or object selected in the Navigator window.

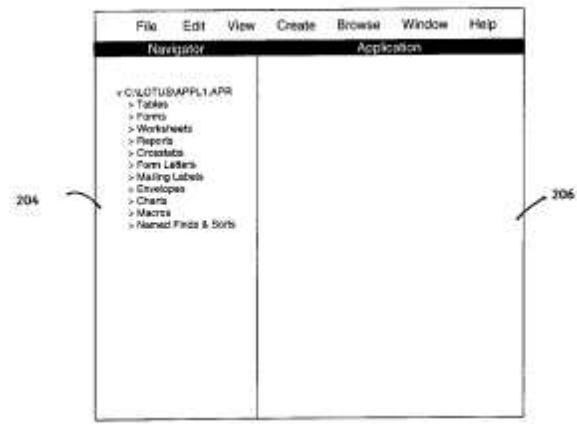

**TRIZ based analysis**

(The tree navigator is known to the prior art. The invention discloses to use that mechanism with an application program).

Invention displays a navigator window for an application program to display the hierarchy of application objects, such as forms, tables, worksheets, reports, macros etc. to easily expand and collapse the lists (Principle-1: Segmentation).

## 2.2 Large tree structure visualization and display system (6104400)

**Background problem**

Large organizational tree structures are difficult to be visualized by means of regular graphical tree layouts because the graph soon becomes too large due to the overloading of information on the screen. For example, the Windows Explorer displays folders in a tree through a scrollbar mechanism. But such a system is line-based and lacks aesthetic quality. There is a need for a convenient display of large graphical trees with arbitrary number of branches within limited screen size.

**Solution provided by the invention**

Patent 6104400 (invented by Halachmi et al., assignee IBM, issued Aug 2000) provides a solution to visualize a large tree structure having n levels of descendants from the center objects that enables to view the descendants to be viewed on the screen together with the center object.

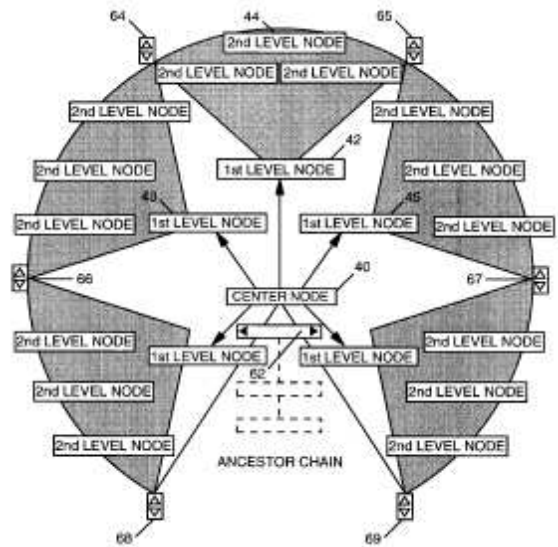

A scrollbar associated with the parent object enables all the children of a parent object to be scrolled in a fan. Fan control buttons on the screen enable the aperture of the fans to be modified so that a variable number of children can be visualized in each fan.

**TRIZ based analysis**

While the conventional tree structure is linear, the invention discloses a fan structure that spreads to all direction (Principle-14: Curvature).

The fan controls are associated with fan control buttons to modify the aperture of the fans centered around the parent object. By using these buttons the user can make one fan bigger while making the other smaller (Principle-15: Dynamize).

## 2.3 Graphical user interface inline scroll control (6181316)

**Background problem**

The window based user interfaces (such as tree navigators) use scrollbars to move large workspaces up and down in a small window. Alhough a scrollbar is a very useful control on the GUI, it consumes fixed amount of space and remains on screen through out the existence of the window. The other disadvantage of a scrollbar is that it requires long pointer movements as it is placed at the side and bottom of the window. There is a need to save the screen space typically occupied by scrollbars and reduce the pointer movements to operate the scrollbars.

**Solution provided by the invention**

Little et al. invented an inline scroll control (Patent 6181316, assigned to IBM, Jan 2001) which places the scroll control directly on the worksheet. The scrolling is achieved by clicking on the up and down indicators. This inline scrolling is designed to reduce the amount of space required on a display device to convey information to the user.

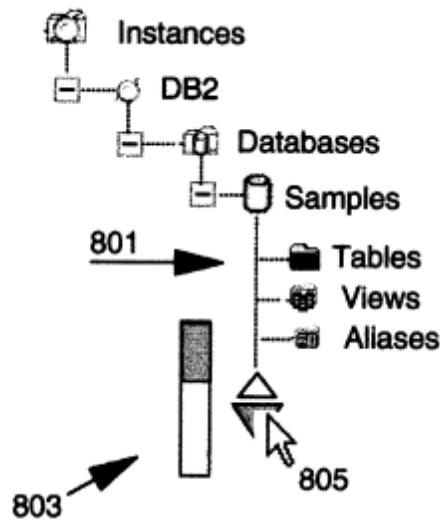

The scrollbars are small in size (like buttons), which requires less pointer movements. Reducing pointer movement is very useful in portable computers having trackballs.

**TRIZ based analysis**

The invention displays the scrollbar on the worksheet itself instead of displaying at the side and bottom of the window (Principle-17: Another Dimension).

The size of the scrollbar is reduced to reduce the pointer movements (Principle-35: Change parameter).

## 2.4 Programmable tree viewer graphical user interface with integrated control panel (6348935)

**Background problem**

A treeview is used to display and mange hierarchical data. Existing treeview controls have very limited capabilities for performing manipulating hierarchical data and customising the display. There is a need for an improved GUI for the treeview.

**Solution provided by the invention**

Malacinki et al. invented a composite object (patent 6348935, assigned by IBM, issued Feb 2002) which group together with tree viewer object and control function for manipulating the format and contents of the tree viewer object in a single object. This new user interface object simplifies manipulation of hierarchical data displayed in its associated treeview display.

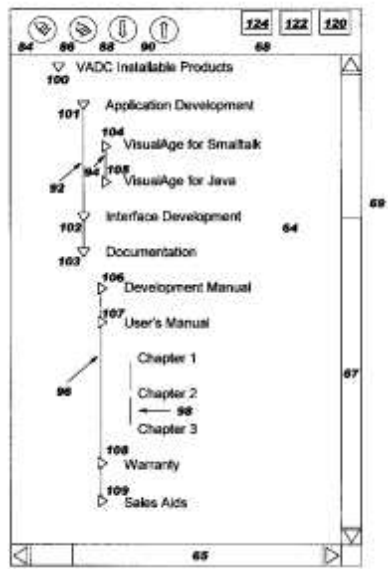

This composite graphical user interface includes a display window for the treeview display and a treeview control panel displayed adjacent to the treeview display. The treeview display and the treeview control panel are displayed within the display window so as they appear as a single integrated graphical user interface.

**TRIZ based analysis**

The invention integrates the treeview control function in the treeview display object (Principle-40: Composite).

The user can customize the treeview by choosing treeview properties from a predefined set of parameters (Principle-15: Dynamize).

## 2.5 Asynchronous tree navigator graphical user interface and associated methods (6559871)

**Background problem**

A tree representation is generally used to present hierarchical information. The data is presented in branches and leaves where the user can "expand" and "collapse" the branches of the tree.

When the hierarchy of the tree and number of items increases a tree navigator becomes complex. In some cases where the data to be retrieve through queries from remote servers, the construction of a tree navigator may take significant amount of time which results in user frustration and confusion. There is a need to improve the tree navigator graphical user interface.

**Solution provided by the invention**

Patent 6559871 (invented by Brozowski et al., assigned by IBM, issued May 2003) discloses a tree navigator having improved responsiveness. According to the invention the graphical tree-navigator is presented asynchronously when the data is being queried from the server. This feature allows the user to start operations on the tree without waiting for the complete data is fetched.

The invention makes the data available on the tree as soon as it is available. Besides the tree navigator uses "placeholder objects" and "loading icons" to notify the user regarding the loading status of the data for a particular branch of the tree.

**TRIZ based analysis**

The invention executes asynchronous queries and displays the tree even before all the data is available to be loaded on to the tree (Principle-10: Prior action).

The invention allows user to work on a loaded branch of the tree while the data is being queried for other branches in a slow connection. The data is made available on the other branches as soon as they are fetched from the database (Principle-20: Continuous action).

The tree navigator becomes available immediately with partially loaded data instead of waiting for the complete data to be loaded (Principle-16: Partial or excessive action).

The navigator uses "placeholder objects" and "loading icons" to inform users about the loading status of the data so that the user is not frustrated (Principle-8: Counterweight).

The "placeholder objects" are displayed in the tree for a temporary period when the actual data is being fetched the branch. The actual objects replace such "placeholder" objects as soon as they are available (Principle-27: Cheap and disposable).

## 3. Summary

The above-illustrated inventions show various improvements in the conventional tree navigator interface. There is scope for further improvements of a tree navigator that we will discuss in separate articles. We can expect to see more and more patents on this topic in near future.

### Reference:

1. US Patent 6078327, "Navigating applications and objects in a graphical user interface", invented by Liman et al., assigned by IBM, issued Jun 2000.

2. US Patent 6104400, "Large tree structure visualization and display system", invented by Halachmi et al., assignee IBM, issued Aug 2000.

3. US Patent 6181316, "Graphical user interface inline scroll control", invented by Little et al., assigned by IBM, issued Jan 2001

4. US Patent 6348935, "Programmable tree viewer graphical user interface with integrated control panel", invented by Malacinki et al., assigned by IBM, Feb 2002.

5. US Patent 6559871, "Asynchronous tree navigator graphical user interface and associated methods", invented by Brozowski et al., assigned by IBM, issued May 2003.

6. Umakant Mishra, "10 Inventions on Scrolling and Scrollbars in Graphical User Interface", July 2005, TRIZsite Journal

7. US Patent and Trademark Office (USPTO) site, http://www.uspto.gov/